\documentclass[11pt]{article}
\usepackage[T1]{fontenc}
\usepackage[utf8]{inputenc}
\usepackage{lmodern}
\usepackage[margin=1in]{geometry}
\usepackage{amsmath}
\usepackage{graphicx}
\usepackage{booktabs}
\usepackage{authblk}
\usepackage[font=small,labelfont=bf]{caption}
\usepackage{xurl}
\usepackage[hidelinks]{hyperref}
\hypersetup{pdfauthor={Rohit Sharma, Pavani Ayinampudi, Aditya B.M.V., Jinal Gupta, Prakash Hegade, Sakshi Sharma, Meenakshi V, SRS Iyengar},pdftitle={An Audit of Measurement Quality and Answer Bias in a Large Classroom-Poll Corpus}}

\title{An Audit of Measurement Quality and Answer Bias in a Large Classroom-Poll Corpus}
\author[1]{Rohit Sharma}
\author[2]{Pavani Ayinampudi}
\author[2]{Aditya B.M.V.}
\author[2]{Jinal Gupta}
\author[2]{Prakash Hegade}
\author[1]{Sakshi Sharma}
\author[1]{Meenakshi V}
\author[1]{SRS Iyengar}
\affil[1]{Indian Institute of Technology Ropar, Rupnagar, Punjab, India}
\affil[2]{ANNAM.AI, Rupnagar, Punjab, India}
\date{}

\begin{document}

\maketitle

\begin{abstract}
Real-time classroom polls are widely used and increasingly generated with automated assistance, yet the questions themselves are rarely evaluated as measurements. We audit a large corpus of authentic classroom polls, 604 items across 47 sessions answered 340,668 times by 2,807 learners, as a measurement instrument. For the 539 items whose correct answer could be established and verified from the lecture transcript, we place every item and every student on a common scale using item response theory and analyse the answer structure of the True/False items. Two findings emerge. First, the polls form a coherent but easy scale of moderate precision (marginal reliability about 0.60), on which roughly a quarter of items barely separate stronger from weaker students. Second, students show a robust tendency to answer True, present at the individual level (77\% of students lean True), which meets a milder tendency for items to be keyed False; as a result answer direction predicts difficulty, False-keyed items being about thirteen points harder, and the effect survives controls for item content and for selective answering. Both findings rest on signals a polling system already records, so the same checks can be run as items are generated, before they reach students.

\end{abstract}

\section{Introduction}

Classroom polls are often interpreted as direct measures of student learning, with the proportion of correct responses treated as an indicator of the extent to which students have understood the instructional content. The question underlying such a measure is frequently designed with limited consideration of its properties as an assessment instrument. Despite this limitation, real-time polling has become increasingly prevalent in contemporary teaching environments. Classroom response systems, web-based polling platforms, and polling features integrated into widely used videoconferencing tools enable instructors to introduce questions during instruction and obtain student responses within seconds~\cite{caldwell2007,kay2009,hunsu2016}. The pedagogical rationale for this practice is well established. Retrieval practice, even under low-stakes conditions, can strengthen memory more effectively than repeated study~\cite{roediger2006,karpicke2008}, while strategically embedded questions can help mitigate the decline in learner attention that commonly occurs during extended lectures~\cite{freeman2014}. Consequently, a live poll serves a dual function. It constitutes a pedagogical intervention that promotes active engagement and retrieval while simultaneously generating data that may be interpreted as evidence of student learning.

When a poll is treated as an assessment instrument, it rarely meets the standards expected of a purpose-designed item. Such an item is expected to have a known level of difficulty, to distinguish between students who have and have not mastered the material, and, in the case of multiple-choice questions, to present distractors that are plausible without being misleading~\cite{haladyna2002,gierl2017}. In practice, relatively little is known about whether individual poll items actually measure what they are intended to, whether they carry a systematic bias, or whether they can be answered correctly without a genuine understanding of the material. This concern has grown in recent years as automated tools are increasingly used to create such items from lecture material~\cite{elkins2024,kurdi2020}. Because these tools are also known to produce fluent, confident output that may be incorrect~\cite{ji2023}, the lack of a measurement perspective is a practical problem rather than a purely theoretical one.

We address this gap by treating a large corpus of authentic classroom polls as a measurement instrument and assessing how well it performs as one. Our corpus comprises 604 poll items collected across 47 sessions of an online summer internship orientation programme, with 340,668 responses from 2,807 learners, as described in Section~\ref{sec:data}. We established and independently verified correct answers from the lecture transcript, without using the recorded votes to determine the key. The subsequent modelling is therefore confined to the 539 items for which the transcript settles the answer. Our contribution is a psychometric and answer-structure audit of a live poll corpus at scale. The calibration separates item-level measurement properties from differences in student performance and identifies quality deficiencies that raw scores can conceal. The answer-structure analysis provides evidence of an individual-level acquiescence tendency whose interaction with item keying makes answer direction predictive of item difficulty. Because the signals underlying both findings, including discrimination, measurement precision, and answer-direction balance, can be computed from response logs alone, the Discussion sets out what a system that generates classroom questions could check before an item reaches students.

\section{Background Study}
\label{sec:rw}

An audit of this kind draws on three bodies of work, namely research on classroom response systems through which these polls are delivered, the psychometric tradition that defines what constitutes sound measurement, and the study of systematic response bias. The first has been investigated for approximately two decades, with generally favourable findings. Reviews of the literature~\cite{fies2006,caldwell2007,kay2009} document gains in participation, attention, and the timeliness of feedback, while a meta-analysis covering over fifty separate studies~\cite{hunsu2016} characterises these gains as genuine but modest and contingent on how the system is used. Taken together, these studies suggest that the benefits of polling depend more on the surrounding pedagogy than on the technology itself, consistent with wider findings that active instructional formats can outperform passive lecturing in science and engineering~\cite{hake1998,freeman2014,black1998,black2009,shute2008,roediger2006,karpicke2008}. This literature examines whether polling is beneficial. Whether the individual questions used in polling provide sound measurement is a separate question that it has not addressed.

Item response theory (IRT) is an established approach for placing test items and respondents on a common latent scale~\cite{rasch1960,lord1980,embretson2000,vanderlinden1997,baker2004}. Its principal advantage over raw percent-correct is that it separates item difficulty from student ability rather than confounding the two, quantifies how sharply each item discriminates, and expresses measurement precision as a function of ability. Classical item analysis provides complementary guidance on item construction and, for multiple-choice items, on distractor quality~\cite{haladyna2002,gierl2017}. These methods are routine for designed examinations, but to our knowledge have not been applied to a large corpus of live, informal classroom polls.

A separate tradition examines response sets, which are systematic tendencies to answer in a fixed manner independently of content. The most extensively documented is acquiescence, the disposition to agree or to respond ``true'' or ``yes'' when uncertain~\cite{cronbach1946,cronbach1950,weijters2013}. Acquiescence is consequential for true/false items because it can interact with the direction in which items are keyed, allowing answer direction to influence apparent difficulty. Automatic question generation has also advanced rapidly, from neural generation of reading-comprehension questions to systems that draft quiz items aligned with Bloom's taxonomy~\cite{du2017,wang2018,elkins2024,kurdi2020}. Such systems are predominantly evaluated using surface criteria, such as fluency and answerability, rather than whether a generated item provides sound measurement~\cite{kurdi2020,kasneci2023}. These traditions leave several aspects of live classroom polling unexamined. A large corpus of live, informal classroom polls has not been audited as a measurement instrument, its answer structure has not been characterised at the respondent level, and neither property has been shown to be recoverable from the response logs already collected by such systems.

\section{Methodology and Methods}
\label{sec:method}

\subsection{Research Questions}
The study is organised around a single overarching question: To what extent do live classroom polls function as sound measurement instruments? We decompose this question into three sub-questions.

\begin{list}{}{%
  \setlength{\leftmargin}{2.6em}%
  \setlength{\labelwidth}{2.1em}%
  \setlength{\labelsep}{0.5em}%
  \setlength{\itemsep}{2pt}%
  \setlength{\parsep}{0pt}%
  \setlength{\topsep}{4pt}%
  \renewcommand{\makelabel}[1]{\textbf{#1}\hfil}%
}
\item[RQ1.] Do the polls form a coherent measurement instrument, and with what precision do they measure?
\item[RQ2.] Do the True/False items exhibit a systematic answer bias, and is this bias associated with item difficulty?
\item[RQ3.] Is the calibration robust to students' self-selection of the polls they answer?
\end{list}

RQ1 examines whether the polls measure a coherent underlying trait and establishes the basis for the subsequent analyses. RQ2 examines the answer structure of the items. RQ3 examines whether the calibration remains stable under the self-selection inherent in a live polling setting. Every quantity used to answer these questions is derived from the response logs a polling system already collects, a point taken up in the Discussion.

\subsection{Measurement Model}
We fit three item response models to the response matrix using marginal maximum likelihood with the \texttt{girth} package~\cite{girth2021}. These include the Rasch one-parameter model, a two-parameter model that additionally estimates item discrimination, and a three-parameter model that adds a lower guessing asymptote. We estimate student ability using expected a posteriori (EAP), and we use a synthetic recovery test to verify that the procedure reproduces known item and person parameters before fitting the real data. We select the model for the main analysis using information criteria and nested likelihood-ratio tests (AIC, BIC, and LRTs across the one-, two-, and three-parameter models), with the marginal log-likelihood evaluated on a 61-point normal-quadrature grid. For each item, we report its discrimination with a standard error obtained from item-level logistic information conditional on the EAP abilities. We propagate the standard error of measurement of ability into every student-level claim.

\subsection{Answer Key}
The poll exports record each student's selection but not the correct option. A correct answer was therefore established from the lecture transcript for every poll that has one. For each item, the surrounding lecture was read independently of the recorded votes, the answer the lecture determines was recorded together with a supporting quotation and a confidence rating of high, medium, or low, and the key was verified before use. Verification consisted of a second, independent reading of the transcript for every item on which the class majority disagreed with the derived key (90 items); this reading confirmed the key on 62 and corrected it on 28, and every correction is folded into the key used here. The key was produced by a single reader, so no inter-rater statistic is reported; instead the confidence rating is carried into the analysis. Polls without an objective answer, such as opinion, mood, and preference items, were excluded. The resulting key settles 539 of the 604 polls (89\%), which constitute the modelled set. To guard against uncertainty in the key, every answer-bias result is additionally recomputed on the 353 high-confidence True/False items alone (Section~\ref{sec:results}).

\section{Data Collection and Analysis}
\label{sec:data}

\subsection{Study Context and the Poll System}
Our data come from a summer internship orientation programme conducted entirely online, comprising 47 live sessions delivered across 39 days by several speakers. The programme is intended to introduce incoming interns to the internship and to motivate them, rather than to assess mastery, and its polls accordingly emphasise professional culture and initial exposure to concepts. During each session, a speaker interjects short polls, predominantly in two-option True/False form with a smaller number of multiple-choice items, which students answer live on their own devices. This orientation context is relevant to interpretation because it is consistent with the strong predominance of lower-order questions in the corpus. Throughout, we distinguish findings that appear specific to a setting of this kind from those that may extend more broadly.

\subsection{Data Sources and the Verified Corpus}
The study draws on an anonymised export of the poll logs. Of the 539 items with an established answer key, 517 are two-option items and 22 are multi-option items. One multi-option item is a free-numeric-entry poll whose option count was recorded incorrectly by the ingestion pipeline. We corrected that record and excluded the item from all multiple-choice analyses, leaving 21 genuine multiple-choice items. Of the 517 two-option items, 10 offer options other than True and False, such as Yes/No or Real/Fake, and are set aside, so the answer-bias analysis uses the remaining 507 True/False items. Restricting attention to the 539 items with an established key yields a student-by-item matrix of 2,373 learners, of the 2,807 enrolled, and 539 items. The matrix contains 311,227 binary (correct/incorrect) responses, has a density of 24.3\%, a median of 36 items answered per student, and an overall correct rate of 68\%. Participation is highly uneven. The number of items a student answered has a median of 36 but a mean of 131 (interquartile range 8 to 234), 28\% of students answered fewer than ten items, and the most active fifth of students supplied 66\% of all responses. This is a long tail of light participants rather than attrition: the number of distinct students responding per day rose from about 540 in the first week to about 645 in the final week, since answering polls was expected of students who continued in the programme. Section~\ref{sec:results} examines whether this pattern distorts the calibration. IRT treats the unanswered cells as missing by design, an assumption examined directly in Section~\ref{sec:results}.

\subsection{Item Descriptors and Data Protection}
Two descriptors, both inferred from the surrounding lecture rather than the item wording, are assigned to each of the 539 modelled items. The first is a cognitive level from Bloom's taxonomy~\cite{anderson2001,bloom1956}. The corpus is dominated by lower-order items, chiefly the Knowledge and Comprehension levels. The second is an instructional function, a finer description of the pedagogical role a poll performs, defined using the seven-category scheme shown in Table~\ref{tab:functions}. We treat this scheme as a practical shorthand rather than a definitive taxonomy. These descriptors are used solely as grouping variables and covariates. Characterising the corpus through them is not a contribution of this paper.

\begin{table}[htbp]
\centering
\caption{The seven instructional functions.}
\label{tab:functions}
\small
\begin{tabular}{@{}p{4.2cm}p{10.4cm}@{}}
\toprule
\textbf{Function} & \textbf{Definition} \\
\midrule
Understanding Verification & Checks that an idea just presented has landed \\
Factual Recall & Recall of a specific fact from the lecture \\
Attention/Distortion-Check & Restates lecture content with a planted error, to catch inattention \\
Concept Reinforcement & Reaffirms a stated principle or value \\
Applied Execution & Applies a rule or procedure to a given case \\
Concept Integration & Links two or more ideas from the lecture \\
Other Recall/Comprehension & Lower-order items outside the categories above \\
\bottomrule
\end{tabular}
\end{table}

Table~\ref{tab:examples} shows two representative items with their response splits; on the False-keyed item, answering True is the acquiescent error.

\begin{table}[htbp]
\centering
\caption{Two example True/False polls and their response splits.}
\label{tab:examples}
\small
\begin{tabular}{@{}p{7.4cm}p{3.2cm}lrr@{}}
\toprule
\textbf{Poll (posed as True/False)} & \textbf{Function} & \textbf{Key} & \textbf{True} & \textbf{Correct} \\
\midrule
``A `golden cage' refers to a job with high motivation factors and low hygiene factors.'' & Attention/Distortion-Check & False & 45\% & 55\% \\
``In the vehicle registration example, car number 1389 would be stored on page 89.'' & Applied Execution & True & 90\% & 90\% \\
\bottomrule
\end{tabular}
\end{table}

No raw identifier appears in any result. Names and email addresses are retained in an access-controlled store excluded from version control, and wherever a learner must be tracked across the data, a salted one-way hash of the email address is used in its place. All reported quantities are therefore aggregates over anonymous keys, and no individual-level identifiers are reported. Prior to data collection, informed consent was obtained from all participating students regarding the use of their data for research purposes.

\subsection{Analysis}
The assumptions underlying the latent scale are examined on the binary data directly, rather than through the imputed correlations a naive factor analysis would require. Dimensionality is assessed by parallel analysis of the tetrachoric correlation matrix of a dense, high-response block of items; local independence is assessed by Yen's $Q3$ computed on the model residuals, contrasting within-session with between-session item pairs, since a shared session is the most plausible source of dependence. Convergent validity is evaluated by correlating calibrated difficulty with percent-correct and calibrated ability with raw accuracy.

For the answer-bias analysis, both key direction (the proportion of items keyed False) and response direction (the proportion of True responses) were tested against a chance baseline. Because a two-option item has only two possible responses, each student's True/False choice follows from whether the response was correct and the direction in which the item is keyed, and was reconstructed for every item on that basis. This allowed acquiescence to be examined at the individual learner level rather than only at the aggregate level.

The observed direction effect was then examined against two possible explanations. First, the effect could reflect differences in content difficulty. This possibility was assessed by controlling for instructional function in a student-clustered logistic model of item incorrectness. Second, the effect could result from differences in who responded to the items. This possibility was examined by modelling difficulty as a function of key direction and log response count, and by stratifying the direction analysis by exposure.

\section{Results and Discussion}
\label{sec:results}

\subsection{Results}
\begin{figure}[htbp]
\centering
\includegraphics[width=0.78\textwidth]{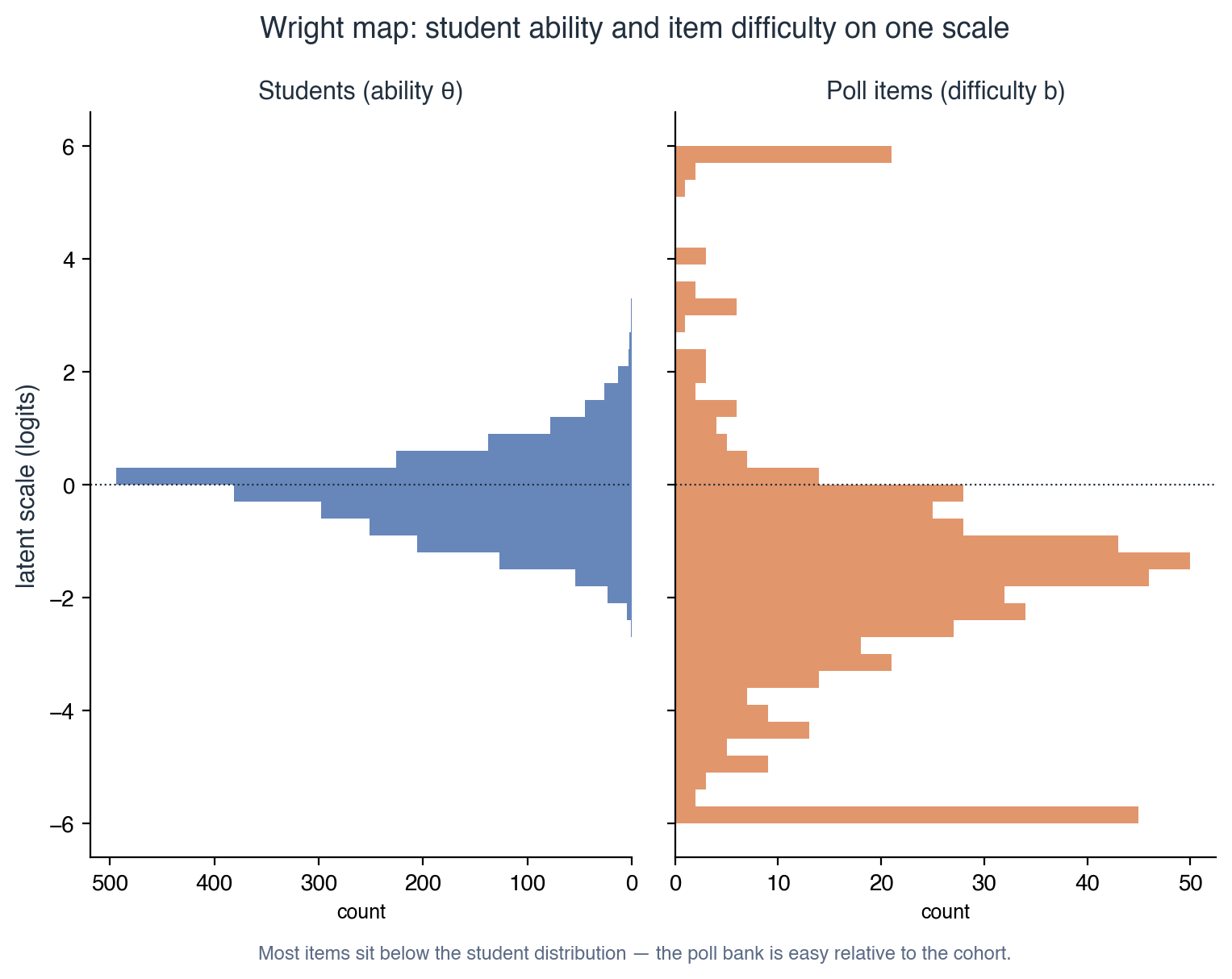}
\caption{Wright map: student ability (left) and item difficulty (right) on one latent scale. The item mass sits below the student mass, indicating that the items are easy for this cohort.}
\label{fig:wright}
\end{figure}

\textbf{The measurement instrument (RQ1).} Considered as a test, the 539 items form a single coherent scale, but with relatively low difficulty. Item difficulty averages $b=-1.59$, placing most of the items below the ability distribution of the students answering them (Fig.~\ref{fig:wright}), consistent with the overall correct rate of 68\%. Model comparison favours the two-parameter model: it minimises both AIC and BIC, the improvement from one parameter to two is decisive ($\chi^2(539)=6579$, $p<10^{-300}$), and the further step to three parameters yields a negligible gain ($p=0.01$) that neither criterion rewards. We therefore adopt the two-parameter model and do not estimate a guessing parameter.

\begin{figure}[htbp]
\centering
\includegraphics[width=0.78\textwidth]{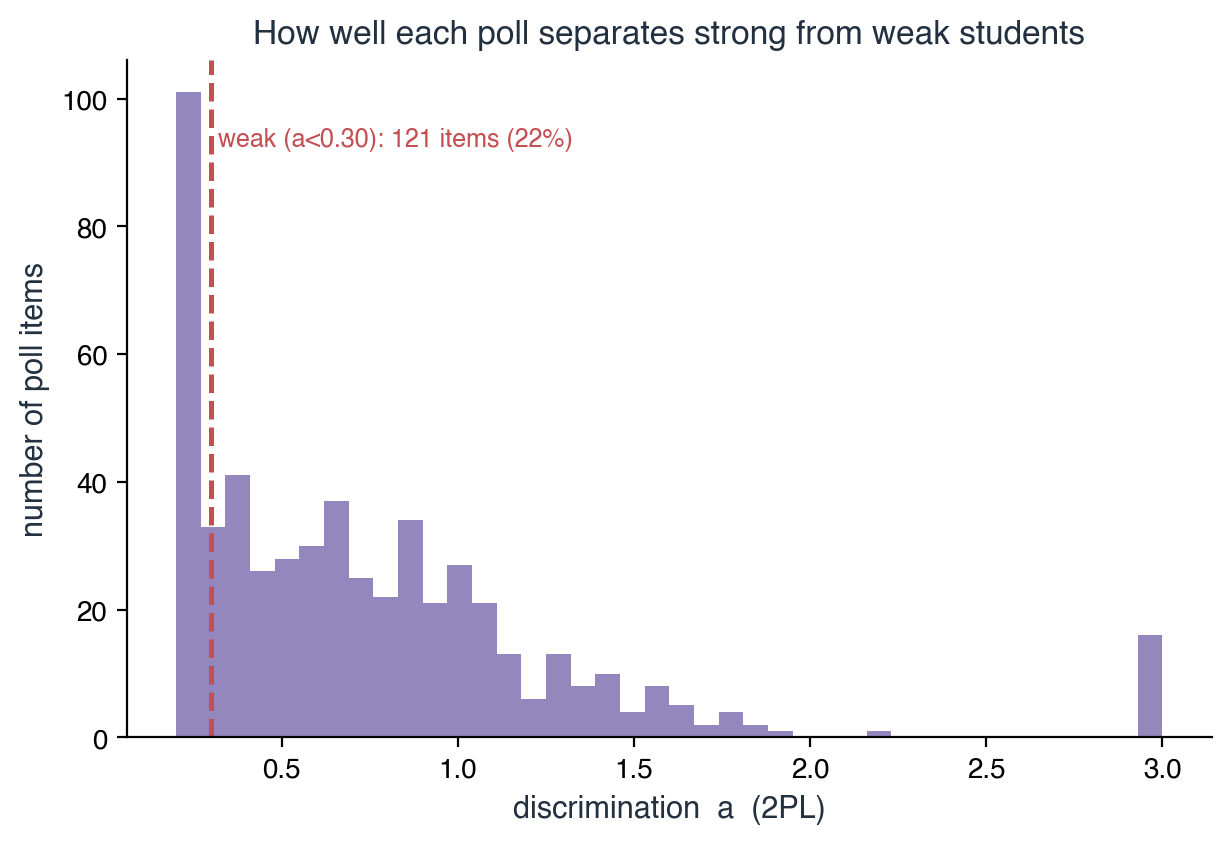}
\caption{Distribution of item discrimination $a$. Nearly a quarter of items fall below the $a=0.30$ weak threshold (dashed).}
\label{fig:disc}
\end{figure}

Discrimination is modest. The median item discrimination is $a=0.65$. By the usual cut-offs, 24\% of items discriminate strongly ($a\geq1.0$), 54\% moderately, and 22\% (121 items) weakly ($a<0.30$; Fig.~\ref{fig:disc}). Because weak discrimination could in principle be an estimation artefact, each estimate carries a standard error (median SE $0.065$). For 65 of the 121 weak items, the upper limit of the 95\% confidence interval remains below the weak threshold, so their weak discrimination cannot be attributed to estimation uncertainty alone. These items are concentrated among routine recall items.

\begin{figure}[htbp]
\centering
\includegraphics[width=0.78\textwidth]{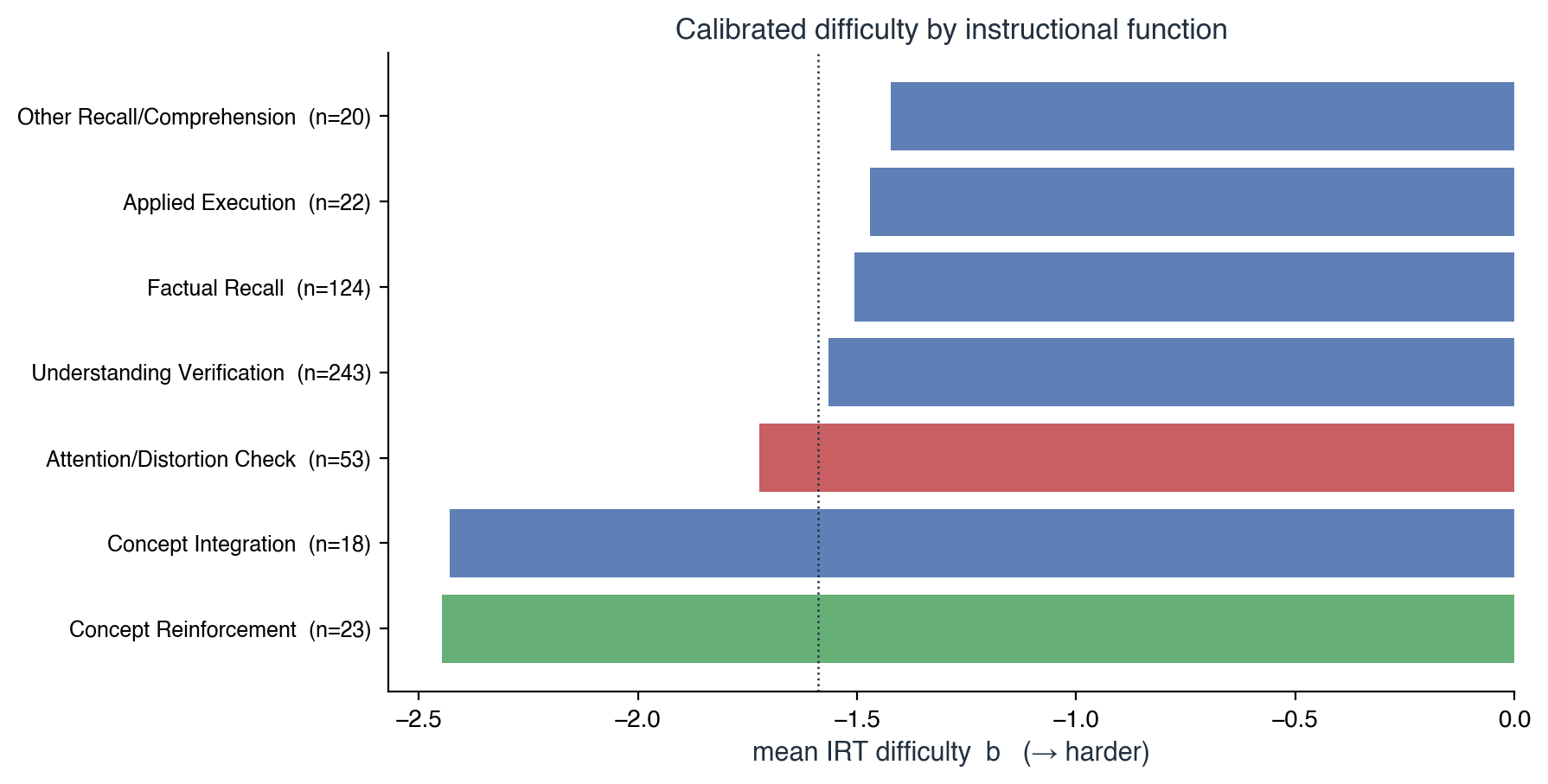}
\caption{Mean calibrated difficulty by instructional function. Concept-Reinforcement and Concept-Integration are the easiest; the remaining functions differ modestly.}
\label{fig:byfunc}
\end{figure}

The scale is approximately unidimensional, and the items show little evidence of local dependence. Item fit is good, with 8 of 539 items exhibiting infit statistics above 1.3. Parallel analysis of the tetrachoric correlations for a dense block of items retains a single dominant factor. Yen's $Q3$ is centred at zero (median $-0.003$), with $|Q3|>0.2$ for only 1.1\% of item pairs, and shows no inflation among within-session pairs. Measurement precision, however, is moderate. The marginal reliability is approximately 0.60, and the median standard error of a student's ability estimate is 0.49 against a latent standard deviation of 0.74. Ability is accordingly interpreted at the level of groups rather than individuals, and a confidence interval accompanies every ability-based comparison. The calibration shows convergent validity, with difficulty correlating $-0.67$ with percent-correct and ability correlating $+0.57$ (95\%~CI $[0.54, 0.60]$) with raw accuracy. Difficulty also increases with cognitive demand (Fig.~\ref{fig:byfunc}). Higher-order Bloom items are harder than Knowledge items, and among the instructional functions, the two reinforcing categories, Concept Reinforcement and Concept Integration, are the easiest by a clear margin. The 21 genuine multiple-choice items are answered correctly at the class level, with the correct option being modal on every item. Thus, confident class-wide error is confined to the True/False portion of the corpus.

\begin{figure}[htbp]
\centering
\includegraphics[width=0.78\textwidth]{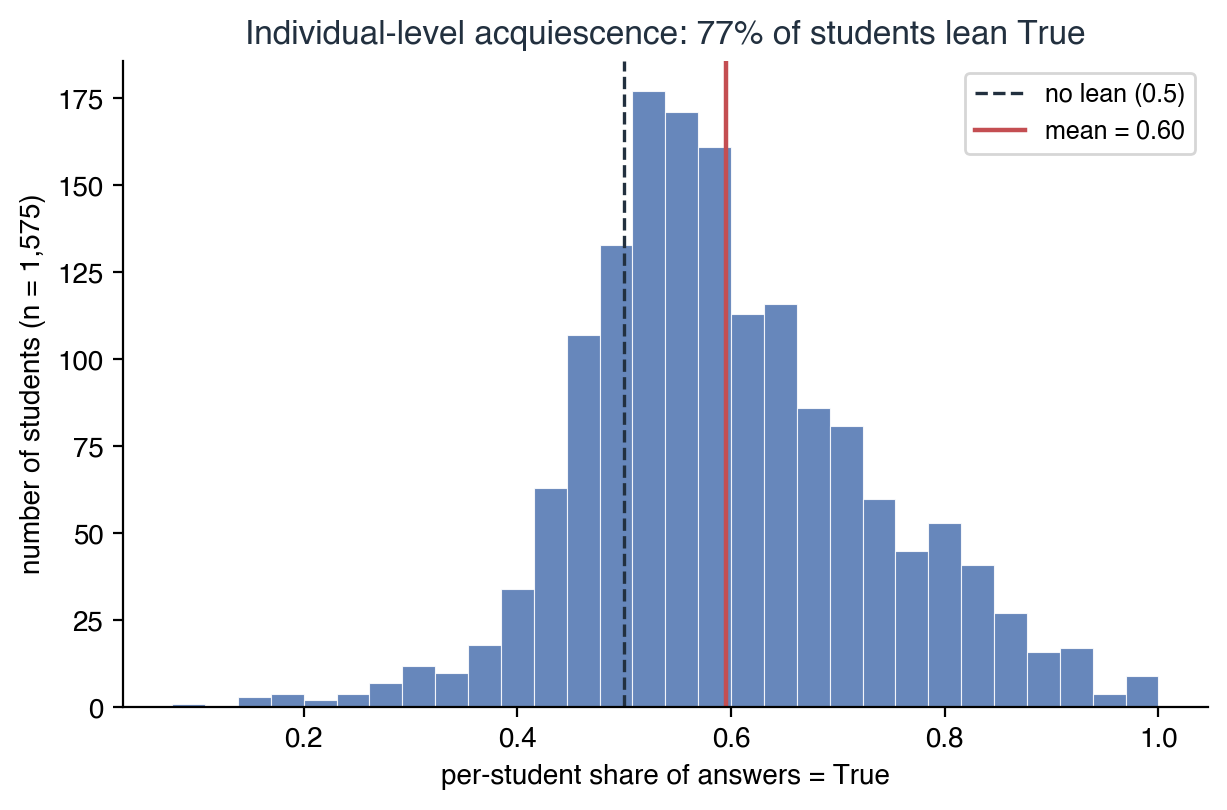}
\caption{Individual-level acquiescence: the distribution of per-student True-rates ($n=1{,}575$ students with at least ten True/False answers) is centred at 0.60, and 77\% of students lean True.}
\label{fig:acq}
\end{figure}

\textbf{Systematic answer bias in the True/False items (RQ2).} The most pronounced answer-structure finding concerns the students' response set. Reconstructing each student's True/False choice reproduces the aggregate lean toward True and provides evidence that the tendency occurs at the individual level. Among the 1{,}575 students who answered at least ten True/False items, the mean individual True-rate is 59.5\%, and more than three-quarters of students lean True (Fig.~\ref{fig:acq}). This pattern is consistent with acquiescence bias.

\begin{figure}[htbp]
\centering
\includegraphics[width=0.78\textwidth]{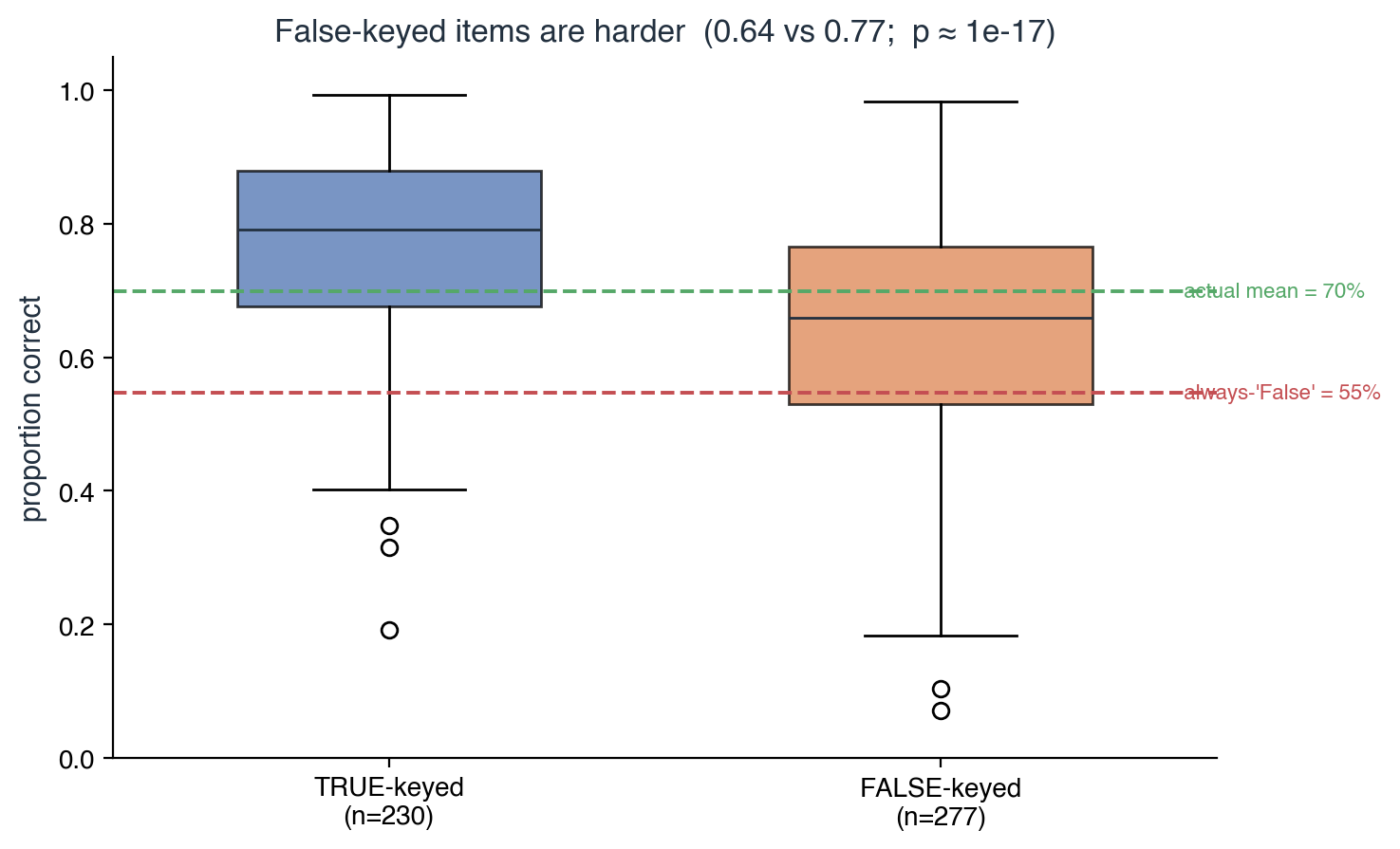}
\caption{Proportion correct by answer direction. True-keyed items are markedly easier than False-keyed items; the always-False and actual-mean baselines are marked.}
\label{fig:dir}
\end{figure}

The item keys lean in the opposite direction, though only mildly. The correct answer is False on 54.6\% of the two-option items, a skew that is marginal over the full set (binomial $p=0.041$) but firmer over the 353 high-confidence items alone (57.8\% False, $p=0.004$). Because students tend to select True while the items are keyed False slightly more often than not, answer direction is associated with difficulty. False-keyed items are markedly harder than True-keyed items, with a mean correct-share of 0.640 against 0.769 (Fig.~\ref{fig:dir}; Mann--Whitney $p\approx10^{-17}$). Two immediate explanations for this pattern can be examined directly. It is not explained by False-keyed items simply posing harder content. After controlling for instructional function in a student-clustered model, a False key raises the odds of an incorrect answer by a factor of 1.79 ($p\approx10^{-102}$). Nor does the pattern appear to be explained by differences in who answered. After controlling for exposure, the False-key penalty on difficulty remains large and positive ($+2.30$ logits, $p\approx10^{-30}$), and the direction gap is nearly identical in the heavily and lightly answered halves of the items ($+0.11$ and $+0.12$). We interpret these results as evidence of an association between the individual response tendency toward True and item keying, while avoiding a causal claim about any individual response.

The interaction has a descriptive consequence and a narrower practical implication. Because the majority key is False, a student answering ``False'' throughout would score 54.6\%, against an actual mean of 69.8\%. This fixed-strategy baseline is reported as a structural property of the item set rather than as evidence that students exploit it, since the polls are ungraded and low-stakes. The gap nevertheless indicates the extent to which answer direction alone can affect scores if the items are reused in a graded or automatically scored setting, making answer-direction balance a check worth performing before such reuse. The skew is also uneven across functions. It is strongest among Applied Execution items, with 74\% keyed False, and Attention/Distortion-Checks, with 58\%, whereas Factual Recall is balanced and Concept Integration leans True.

\begin{figure}[htbp]
\centering
\includegraphics[width=0.78\textwidth]{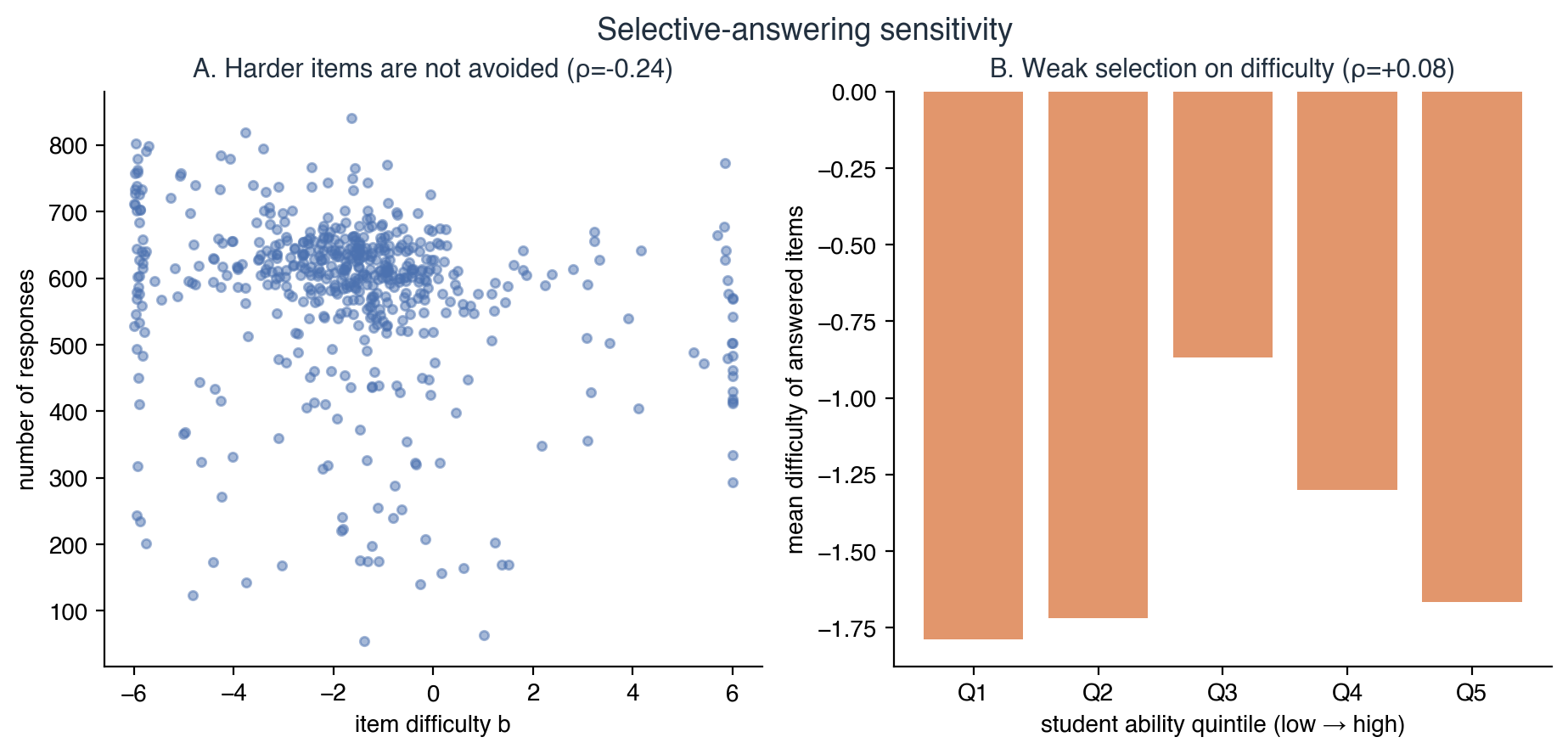}
\caption{Selective-answering probes. (A) Harder items are only weakly under-answered ($\rho=-0.24$). (B) The mean difficulty of the items a student answered is nearly flat across ability quintiles, indicating little selection on difficulty.}
\label{fig:sens}
\end{figure}

\textbf{Robustness to selective answering (RQ3).} Because students answer some polls and not others, the calibration could in principle be affected if answering patterns were associated with item difficulty or student ability. We examine this possibility using four probes (Fig.~\ref{fig:sens}). The principal result is that participation is essentially uncorrelated with estimated ability (Spearman $\rho=+0.03$). This is less surprising in the present setting than it would be elsewhere because answering polls was a requirement for remaining in the programme. Participation therefore reflects attendance rather than the voluntary self-selection that ordinarily governs who responds online. The responding pool also grew rather than shrank over the programme (Section~\ref{sec:data}), so the amount a student answers carries little signal of ability and does not materially alter the scale. The remaining probes give a consistent picture. Harder items are only weakly under-answered ($\rho=-0.24$), there is little selection on difficulty ($\rho=+0.08$, with answered-item difficulty nearly flat across ability quintiles), and refitting the two-parameter model separately on the heavily and lightly participating halves of the cohort yields a difficulty correlation of $\rho=0.70$ over the 143 items with sufficient responses in both. Selective answering therefore appears largely benign for the ability scale. The answer-direction effect reported in the preceding paragraphs also remains after stratification by exposure.

\subsection{Discussion}
The results indicate that the polls measure a coherent construct, but one that is easy, uneven in quality, and, on the True/False items, partly answerable without command of the material. For an instrument whose primary purpose is to hold attention and support engagement, these properties are not necessarily disqualifying, but they constrain its use. The items can provide information about whether the middle and lower portions of a class are keeping pace, but their moderate measurement precision and uneven item discrimination limit their use for ranking students or making higher-stakes inferences. The acquiescence result may have broader relevance because it arises from the interaction between a human response tendency and the direction in which two-option items are keyed, rather than from a feature peculiar to this corpus. A practical implication is to verify answer-direction balance in any two-option item set and, where appropriate, address imbalance through mixed keying or an explicit acquiescence adjustment, particularly when the items are scored.

Calibration adds three things to the simple counts a reader might first examine, such as the share of items keyed False or a student's raw rate of answering True. First, it separates item difficulty from the ability of the particular students who happened to answer each item. This matters when every item is answered by a different subset of a large cohort, because raw correct shares then reflect who answered as well as what was asked. Second, it attaches a standard error to each item's discrimination and each student's ability estimate. This allows us to distinguish the weakly discriminating block from estimation noise and confines ability estimates to group-level use. Third, it provides a model in which the direction effect can be tested while holding content and exposure constant, which is the form in which the bias persists. The imbalance in key direction itself requires no model to detect. Establishing that its association with difficulty is not explained by differences in who answered or in what was asked does.

These observations are useful because the underlying signals are recoverable from data that a polling system already stores. Discrimination and item fit can be estimated from the response matrix, measurement precision from the information function, and answer-direction balance from the item keys alone. None of these checks requires a separate pilot or expert re-rating. A system that authors items automatically could therefore evaluate these signals as part of its item-development workflow, flagging a weakly discriminating item or an imbalance in key direction before the item reaches students. This would extend the retrospective audit reported here into a routine quality check that complements the human review on which such systems still depend~\cite{elkins2024,kurdi2020,ji2023}. The specific magnitudes reported here belong to a single programme and should not be generalised beyond this setting. The methods and the classes of signal they surface may, however, be applicable to other polling settings.

\section{Conclusion}
\label{sec:conc}

This paper has examined a large corpus of live classroom polls as a measurement instrument, a perspective these polls rarely receive. As an instrument, the items are coherent but easy and only moderately precise, with a justified two-parameter model, an approximately unidimensional scale, and roughly a quarter of items discriminating weakly (RQ1). The True/False items carry a systematic answer bias in which a per-student tendency toward True meets a mild tendency to key items False, so that answer direction is associated with difficulty even after content and exposure are controlled for (RQ2). The calibration also appears largely robust to selective answering by students (RQ3). Because the signals underlying both findings are recoverable from response logs, the same checks can complement human review as such questions come increasingly to be authored automatically.

Three directions follow. A richer response-style model would permit acquiescence and ability to be estimated jointly rather than in sequence. The log-computable checks could be embedded within an authoring and review loop, so that a weakly discriminating item or an imbalance in key direction is detected before an item reaches students. A validated per-option distractor analysis, based on a corrected raw-poll parser, could extend the approach to multiple-choice items. Finally, replication across instructors, subjects, and assessment stakes would help establish which findings are specific to a low-stakes orientation setting and which may generalise to rapid, informal, and increasingly machine-assisted questioning.

Two limitations qualify the account. All correctness rests on the transcript-derived and verified key. Although the answer-bias results hold on its high-confidence subset, items that the transcript could not settle are excluded, limiting the scope of the corpus to questions with answers that can be anchored in the instructional content. The key skew is also mild and only marginal over the full set, while reliability is moderate. We therefore frame the bias around the robust per-student response tendency rather than the magnitude of the key imbalance, and we confine ability estimates to group-level use.

\section*{Acknowledgements}
The per-item cognitive-level and instructional-function labels and the answer key were prepared by the authors as part of a broader study of this corpus.

\end{document}